\documentclass[prb,aps,twocolumn,superscriptaddress,showpacs]{revtex4-1}

\usepackage{graphicx}
\usepackage{epstopdf}

\begin{document}

\title{The effect of the pseudogap on suppressing high energy inelastic neutron scattering in superconducting YBa$_{2}$Cu$_{3}$O$_{6.5}$}

\author{C. Stock}
\affiliation{NIST Center for Neutron Research, 100 Bureau Drive, Gaithersburg, Maryland 20899, USA}
\affiliation{Indiana University Cyclotron Facility, 2401 Milo B. Sampson Lane, Bloomington, Indiana 47404, USA}

\author{R. A. Cowley}
\affiliation{Diamond, Rutherford Appleton Laboratory, Chilton, Didcot, OX11 0QX, UK}
\affiliation{Oxford Physics, Clarendon Laboratory, Parks Road, Oxford, OX1 3PU, UK}

\author{W. J. L. Buyers}
\affiliation{National Research Council, Chalk River, Ontario, K0J 1JO, Canada}
\author{C.D. Frost} \author{J.W. Taylor}
\affiliation{ISIS Facility, Rutherford Appleton Laboratory, Chilton, Didcot, OX11 0QX, UK}

\author{D. Peets}
\affiliation{Physics Department, University of British Columbia, Vancouver, B. C., V6T 2E7, Canada}

\author{R. Liang} \author{D. Bonn} \author{W.N.  Hardy}
\affiliation{Physics Department, University of British Columbia, Vancouver, B. C., V6T 2E7, Canada}

\date{\today}

\begin{abstract}

	We have measured the spin fluctuations in the YBa$_{2}$Cu$_{3}$O$_{6.5}$ (YBCO$_{6.5}$, T$_{c}$=59 K) superconductor at high-energy transfers above $\sim$ 100 meV.  Within experimental error, the momentum dependence is isotropic at high-energies, similar to that measured in the insulator for two dimensional spin waves, and the dispersion extrapolates back to the incommensurate wave vector at the elastic position.  This result contrasts with previous expectations based on measurements around 50 meV which were suggestive of a softening of the spin-wave velocity with increased hole doping.  Unlike the insulator, we observe a significant reduction in the intensity of the spin excitations for energy transfers above $\sim$ 100 meV similar to that observed above $\sim$ 200 meV in the YBCO$_{6.35}$ (T$_{c}$=18 K) superconductor as the spin waves approach the zone boundary.  We attribute this high energy scale with a second gap and find agreement with measurements of the pseudogap in the cuprates associated with electronic anomalies along the antinodal positions.  In addition, we observe a sharp peak at around 400 meV whose energy softens with increased hole doping.  We discuss possible origins of this excitation including a hydrogen related molecular excitation and a transition of electronic states between $d$ levels.

\end{abstract}


\maketitle

\section{Introduction}

	The behavior of the magnetic fluctuations of layered cuprates is strongly coupled to their superconductivity. The  parent compounds are antiferromagnetic Mott insulators.~\cite{Stock06:75,Kastner98:70} The materials are then superconducting if the CuO$_{2}$ planes are doped with holes whose concentration exceeds the critical value p$_{c}$=0.055 (Ref. \onlinecite{Fujita02:65}) when the long range antiferromagnetism is largely suppressed. In the hole overdoped and non superconducting region, the superconducting phase is replaced by a Fermi liquid and the magnetic excitations are largely absent.~\cite{Proust02:89,Waki04:92} Despite a significant effort, there is no complete understanding of the nature of the magnetic excitations, or of their role in the superconductivity or their hole doping dependence.  The magnetic excitations can be studied by neutron inelastic scattering techniques but  owing to the large exchange constants resulting in excitations with energies well over $\sim$ 100 meV~\cite{Reznik96:53} the measurements are difficult with reactor sources. Measurements at these high energies can be made using pulsed spallation sources such as the ISIS facility at the Rutherford Appleton Laboratory.
	
	The magnetic excitations of the parent undoped, antiferromagnetic, and insulating compounds have been well studied in both YBCO$_{6.15}$ and La$_{2}$CuO$_{4}$.~\cite{Hayden96:54,Hayden96:76,Coldea01:86}  Both sets of experiments found the magnetic excitations to be dominated by well defined spin waves emanating from the commensurate antiferromagnetic Bragg peak at $\vec{Q}$=(0.5,0.5) with a large velocity of $\hbar c \sim$ 600 meV \AA\ with the top of the dispersion band located at $\sim$ 300-350 meV.  If the hole concentration is only slightly above the critical concentration, the magnetic scattering consists of short-range magnetic order and at low energies slow excitations characteristic of ordered magnetic regions separated by charge rich regions as we observed forYBCO$_{6.35}$ (T$_{c}$=18 K).~\cite{Stock06:73,Stock08:77} 

	For hole concentrations within the superconducting phase of YBCO, the low energy commensurate spin fluctuations are replaced by incommensurate rods of scattering which disperse towards a commensurate resonance peak at a finite energy transfer.~\cite{Li08:77,Hinkov08:319}  The low-energy incommensurate scattering is highly anisotropic with peaks only displaced along one reciprocal lattice direction.~\cite{Waki00:61,Mook00:404}  The value of the incommensurate displacement scales with hole doping and with superconducting transition temperature.~\cite{Yamada98:57,Balatsky99:82}  Both the hole doped YBCO$_{6+x}$ and La$_{2-x}$Sr$_{x}$CuO$_{4}$ systems give the same values for the incommensurability for a given hole doping $p$.~\cite{Dai01:63}  

	The resonance peak defines a crossover energy that separates low-energy incommensurate fluctuations from high-energy fluctuations which have more symmetric line shapes in momentum than the low-energy incommensurate fluctuations.~\cite{Dai99:284,Arai99:83}  The resonance peak in neutron scattering appears as a commensurate peak in momentum and has a well-defined lineshape in energy as measured by a constant-$Q$ scan at the antiferromagnetic zone center.~\cite{Fong00:61}  While there are theoretical and experimental studies relating the resonance energy to superconductivity (Ref. \onlinecite{Onu09:102,Yu09:5,Sidis04:6}), the resonance energy does not universally scale with T$_{c}$ or the gap value where it has been observed, most notably in YBCO$_{6+x}$ (Ref. \onlinecite{Fong00:61}), La$_{2-x}$Sr$_{x}$CuO$_{4}$, Tl$_{2}$Ba$_{2}$CuO$_{6+\delta}$ (Ref. \onlinecite{He02:295}), Bi$_{2}$Sr$_{2}$CaCu$_{2}$O$_{8+\delta}$ (Ref. \onlinecite{Fong99:398}), and HgBa$_{2}$CuO$_{4}$ (Ref. \onlinecite{Yu08:xx}).  For example,  for  YBCO$_{6.5}$ the energy of the resonance is $\sim$ 33 meV while the superconducting temperature is 59 K and for La$_{1.875}$Ba$_{0.125}$CuO$_{4}$ the resonance energy is found to be $\sim$ 60 meV despite a superconducting transition temperature of less than 6 K.~\cite{Tranquada04:429,Xu07:76}   This is further demonstrated by our measurements on YBCO$_{6.35}$ (T$_{c}$=18 K) where no underdamped resonance peak was observed.~\cite{Stock08:77}   There has also been some debate on whether the intensity of the resonance peak is sufficient to account for the electronic anomalies.~\cite{Kee02:88,Abanov02:89}  While there is significant experimental work correlating the resonance energy with the superconducting gap, these studies have typically focussed on a narrow range of hole dopings near optimal hole concentrations. ~\cite{Sidis04:6}   We therefore conclude, based on these examples, that the resonance energy cannot represent a universal magnetic energy scale connected with superconductivity in the cuprates.
	
	The momentum and energy dependence of the magnetic excitations at energies higher than the resonance has also been a topic of some debate with reports suggesting that the momentum dependence resembles squares (Refs. \onlinecite{Hayden04:429,Tranquada04:429}) and others reporting circular patterns (Ref. \onlinecite{Stock05:71}) consistent with spin-wave type  of excitations.  The energy dependence is also not well understood with some groups suggesting near vertical dispersion (Ref. \onlinecite{Hayden04:429}) and others suggesting nearly linear dispersion of the spin excitations.  This behavior of the high-energy magnetic fluctuations is discussed in this paper.

	We have previously investigated the spin fluctuations in samples of YBCO$_{6.35}$ (T$_{c}$=18 K) and Ortho-II YBCO$_{6.5}$ (T$_{c}$=59 K).~\cite{Stock08:77,Stock04:69}  The magnetic excitation spectrum has been mapped in very underdoped yet superconducting YBCO$_{6.35}$ (T$_{c}$=18 K) over the entire band of magnetic excitations.~\cite{Stock07:75}  The low-energy fluctuations result from short-range magnetic order and the slow magnetic fluctuations are characteristic of regions of ordered spins separated by charge rich regions.~\cite{Stock06:73,Stock09:79}  The magnetic fluctuations disperse in a similar manner to the antiferromagnetic insulator with commensurate spin-waves extending up to $\sim$ 200 meV.  At these high-energies, the magnetic excitations become heavily damped in energy, display substantial spectral weight loss, and have a lower energy than that observed in insulating cuprates and predicted from linear spin-wave theory.  We proposed that this was due to the interaction of the magnetic excitations with particle-hole pairs above the so-called pseudogap energy,  for it matched the energy scales observed in photo-emission and thermal conductivity experiments.~\cite{Sutherland03:67,White96:54,Loeser97:56,Norman98:392,Campuzano99:83}  The pseudogap energy was found from the optical conductivity to decrease with increasing hole doping but its origin is not understood.~\cite{Timusk99:62,Stajic03:68}   In this paper we show that the suppression of the neutron scattering cross section seen with neutrons also decreases with increased hole concentration.
	
	We also made earlier measurements of the low-temperature magnetic response of superconducting Ortho-II YBCO$_{6.5}$ up to $\sim$ 100 meV.  At energies above the resonance peak of 33 meV, the anisotropic low-energy incommensurate magnetic response is replaced by isotropic rings of scattering similar to the spin-wave response observed in the heavily underdoped superconducting YBCO$_{6.35}$ compound, but with a much reduced velocity.~\cite{Stock05:71}  The integrated intensity above the resonance energy agreed with linear spin-wave theory for two-dimensional magnetic fluctuations up to about 100 meV where the error bars in the measurements became large.   

	Motivated by the high-energy results in YBCO$_{6.35}$, in this paper, we have extended our investigation of the magnetic fluctuations in Ortho-II YBCO$_{6.5}$ to higher energies.  We will show that the momentum dependence is isotropic in momentum at energy transfers above 100 meV and, in contrast to our previous analysis the velocity is larger and the excitations may originate from the incommensurate points. Furthermore we shall show that the intensity of the excitations shows a marked decrease at energy transfers of $\sim$ 125 meV, an energy scale which agrees with the pseudogap energy for this particular hole concentration.  
	
	Based on these results, we propose that the magnetic excitation spectrum of all cuprates are described by two energy scales, one representing the effects of the superconducting gap and the other the pseudogap.  A detailed comparison with other techniques will be presented.

	In addition to these results, we also present data from very high-energies beyond the top of the one-magnon band in the cuprates and find a mode with an energy of $\sim$ 400 meV.  Its energy scales with the hole doping concentration and the momentum dependence is consistent with a magnetic origin.  Various possible explanations for this peak are discussed including a direct orbital transition.

	The paper is divided into three main sections.  The next section describes the experimental arrangement, the methods and the samples used.  The following section describes the experimental results including a discussion of the momentum dependence, the dispersion and integrated intensity of the magnetic excitations.  This section and an Appendix also discusses the mode observed at high energies near 400 meV. The final section consists of the summary and conclusions. 

\section{Experiment}

	We performed two different sets of experiments using the MAPS and MARI direct geometry time-of-flight spectrometers at the ISIS facility of the Rutherford- Appleton laboratory.  The MAPS position-sensitive-detectors are located 6 m from the sample and are designed so that there is no direction that has significantly poorer ${\bf{Q}}$ resolution arising from the shape of the detectors.  The incident beam was obtained via a Fermi chopper spinning at a frequency of 400 Hz with incident energies fixed to E$_{i}$=400 and 600 meV.  The choice of these energies was  based upon our experimental work in YBCO$_{6.35}$ (T$_{c}$=18 K) where the top of the magnetic scattering lay in the range of 200-300 meV.   The sample was mounted with the [001] axis parallel to the incident beam in a closed cycle refrigerator.  As described in detail in Ref. \onlinecite{Stock05:71}, the value of $L$, and hence the bilayer structure factor, varies with energy transfer.  By varying the angle of the [001] axis of the sample with respect to the incident beam, we showed that the optic and acoustic scattering intensities are equal at energy transfers above $\sim$ 50-60 meV in both YBCO$_{6.5}$ and YBCO$_{6.35}$.  Since we are interested in excitations only above this energy we have not separated the acoustic and the optical contributions as we have done before for the low energy magnetic scattering around and below the resonance energy.  Our results are independent of the value of $L$ and we shall discuss the data only in terms of $H$ and $K$.  Since the [001] axis was parallel to the incident neutron beam, symmetry enables us to gain count rate by adding the four detector banks present at low-angles on the MAPS spectrometer.

	The experimental arrangement is similar to our previous reports on YBCO$_{6.5}$ with E$_{i}$=150 meV.  The goal of this experiment was to measure to higher energies using E$_{i}$=400 meV and E$_{i}$=600 meV so that we could study how the magnetic excitations evolve close to the top of the one-magnon band.

	The spectrometer was calibrated by using a standard vanadium sample of known mass.  The cross section was assumed to be isotropic in spin while the definitions of $\chi''(q,\omega)$ and the total cross-section are the same as we used previously and agree with those of most other groups.  The definitions of the magnetic cross section used by us are compared with those of other groups in Refs. \onlinecite{Stock04:69,Stock05:71}.  We note that the vanadium calibration has been found to be in good agreement using an internal phonon standard. 

	Motivated by the preliminary high-energy results obtained at the MAPS spectrometer, we continued investigations of the high-energy fluctuations using the MARI direct geometry time-of-flight spectrometer.  MARI is particularly useful for studying the momentum dependence of an excitation as the detector bank extends to angles up to $2\theta$ $\sim$ 130$^{\circ}$ whereas the MAPS detector bank only extends to 60$^{\circ}$ in scattering angle. This difference is particularly helpful to distinguish the magnetic scattering from the scattering by the molecular modes involving hydrogen.~\cite{Stock10:81}  The MARI detectors are 30 cm long and 2.5 cm wide and located 4 m from the sample position and the measurements were conducted with the sample in vacuum without the presence of a cryostat or closed cycle refrigerator with the entire spectrometer at room temperature. This eliminated the possibility of observing molecular modes from water condensed on the walls of the cooling system.  For all the measurements we used the relaxed Fermi chopper to provide a monochromatic incident energy of 750 meV while spinning the chopper at 600 Hz.   The sample was mounted with the [001] axis parallel to the incident beam.

	Both the MAPS and MARI measurements were made on a sample of Ortho-II YBCO$_{6.5}$.  The sample consisted of six orthorhombic crystals with a total volume of $\sim$ 6 cm$^{3}$ aligned on a multi-crystal mount with a combined rocking curve width of $\sim$ 2$^{\circ}$.  The samples were aligned such that reciprocal lattice reflections of the form (HHL) lay in the horizontal scattering plane.  Details of the crystal growth, the detwinning by stress along the $a$ axis, and the oxygen ordering have been published earlier.~\cite{Peets02:15}  The sample is partially detwinned with the majority domain occupying 70 \% of the total volume.  The oxygen ordering had a correlation length of $\sim$ 100 \AA\ in both the $a$ and $b$ directions.  The DC magnetization shows that the superconducting onset temperature was 59 K with a width of $\sim$ 2.5 K.  The low temperature lattice constants were $a$=3.81 \AA, $b$=3.86 \AA, and $c$=11.67 \AA. These crystals were used for our previously published studies.~\cite{Stock02:66}  Using MARI, we also studied a multi-crystal array of four $\sim$ 1 cc crystals of YBCO$_{6.33}$ (T$_{c}$=8 K) also coaligned in the (HHL) scattering plane. The YBCO$_{6.33}$ materials were orthorhombic with lattice constants $a$=3.844 \AA, $b$=3.870 \AA, and $c$=11.791 \AA.  The samples were twinned and no oxygen ordering Bragg peaks were observed with neutron scattering techniques.  Since x rays showed orthorhombic symmetry over much shorter distances, only short oxygen chains were present.

\section{Results}

\begin{figure}[t]
\includegraphics[width=8.3cm] {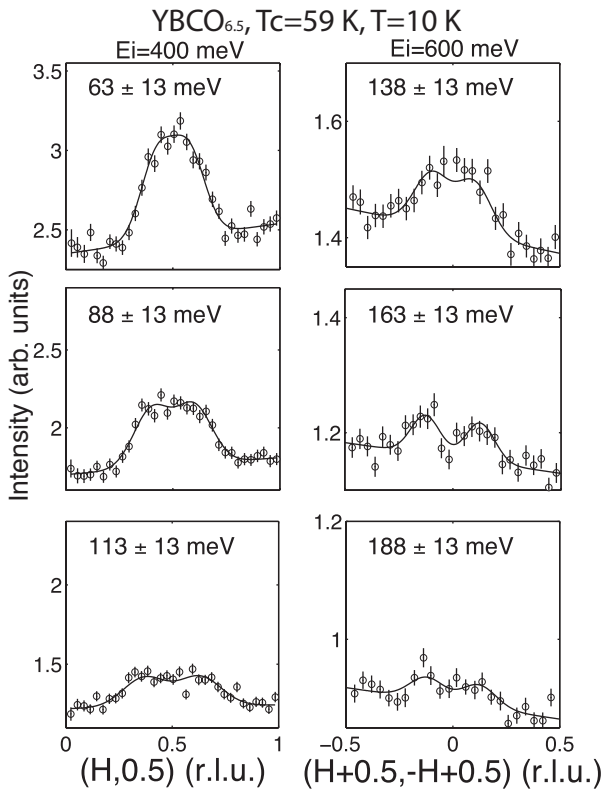}
\caption{\label{1dcuts}  Constant energy cuts through the correlated magnetic scattering with E$_{i}$=400 meV (left hand panels) and 600 meV (right hand panels) obtained with the MAPS direct geometry spectrometer.  The cut directions were chosen to be [1 0] for E$_{i}$=400 meV and [1$\overline{1}$] for E$_{i}$=600 meV.  The E$_{i}$=400 meV data is integrated along K by $\pm$ 0.1 r.l.u. and the E$_{i}$=600 meV is integrated along [-K,K] by $\pm$ 0.05 r.l.u.}
\end{figure}

	In this section, we describe the results from our experiments at ISIS for the samples of YBCO$_{6.5}$ (T$_{c}$=59 K) and YBCO$_{6.33}$ (T$_{c}$=8 K).  The first three subsections focus on the Ortho-II sample of YBCO$_{6.5}$ and describe our extended measurements for energy transfers up to $\sim$ 200 meV.  We will discuss the dispersion of the magnetic excitations, the lineshape in momentum, and the intensity.  In the final section we present our observation of a mode at $\sim$ 370 meV in YBCO$_{6.5}$ and at  $\sim$ 430 meV in YBCO$_{6.33}$ where it illustrates that the energy of this mode scales with the doping.  

\begin{figure}[t]
\includegraphics[width=8.6cm] {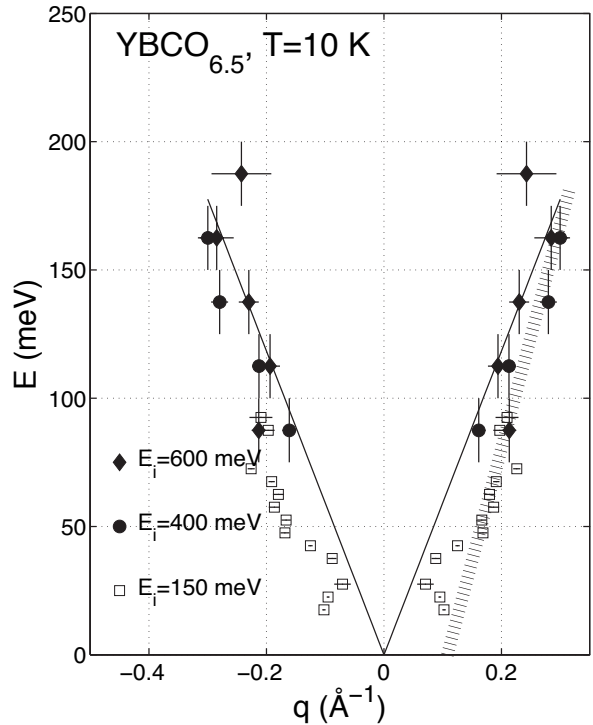}
\caption{\label{dispersion}  The peak positions along the [1 0] direction, with respect to the (H,K)=(0.5, 0.5) position, of the magnetic response as a function of energy transfer for E$_{i}$=400 meV and 600 meV.  Previously published results taken with E$_{i}$=150 meV are plotted for comparison illustrating the consistency of the data and the low-energy response.  The solid line is the spin-wave velocity measured in YBCO$_{6.35}$ of $\hbar c$=590 meV \AA.  The error bar in the velocity is $\pm$ 30 meV \AA.  The dashed line originating from the incommensurate elastic position is discussed in the text.}
\end{figure}

\subsection{Dispersion - High velocity spin waves originating from incommensurate positions}

	We performed constant energy cuts through the MAPS data taken with E$_{i}$=400 meV and 600 meV to extract the momentum dependence of the magnetic fluctuations as a function of the energy transfer.  Representative cuts are presented in Fig. \ref{1dcuts} with cuts taken along the [1 0] direction for E$_{i}$=400 meV and [1 -1] direction for E$_{i}$=600 meV.   The E$_{i}$=400 meV constant energy cuts shown by the left-hand panel capture the overall qualitative momentum dependence of the magnetic fluctuations.  At an energy transfer of $\hbar \omega$=63 $\pm$ 13 meV, the correlated magnetic scattering near the antiferromagnetic wave vector position is well defined and shows little evidence for the presence of two displaced peaks.  At higher energy transfers of 88 $\pm$ 13 meV and 113 $\pm$ 13 meV, the correlated magnetic scattering broadens considerably and develops a clear flat-top structure indicative of the dispersion of the magnetic excitations.  The broadening and splitting becomes even more obvious at the higher energy transfers probed using E$_{i}$=600 meV and in particular for $\hbar \omega \sim$ 163 $\pm$ 13 meV two distinct peaks are observed.  The results in Fig. \ref{1dcuts} clearly show evidence for the dispersion of the magnetic excitations at high-energies above the resonance.  This finding differs from some experiments on YBCO$_{6.6}$ where vertical rods of scattering were proposed to exist above the resonance peak, and were ascribed to a Fermi surface explanation for the magnetic excitations.~\cite{Hayden04:429}

	The dispersion of the magnetic excitations was quantified by fitting to two Gaussians symmetrically displaced from the antiferromagnetic position at $\vec{Q}$=(0.5, 0.5).  This simple model describes the data well over a very broad energy range. The background was assumed to be a constant plus a linear term with increasing momentum transfer.  The values of the peak positions in momentum as a function of the energy transfer are illustrated in Fig. \ref{dispersion} with the momentum transfer along the [1 0] direction.  The solid line in Fig. \ref{dispersion} shows the magnetic excitations measured in YBCO$_{6.35}$ (T$_{c}$=18 K) with a slope of $\hbar c$=590 meV \AA.  The errorbar in the present measurement of the velocity is $\pm$ 30 meV \AA.  The experimental data taken with E$_{i}$=400 meV and 600 meV are represented by filled diamonds and circles.  The previous data taken with E$_{i}$=150 meV are plotted as open squares and are in good agreement with the current experimental data set where there is overlap at $\hbar \omega \sim$ 75 meV.  

	The magnetic fluctuations at energies much below the resonance have incommensurate wave vectors displaced along (0.5 $\pm \delta$, 0.5), as described previously.  These incommensurate fluctuations disperse inwards and meet at the resonance at an energy of $\hbar \omega$=33 meV.  Above the resonance energy, the spin fluctuations disperse outwards and were originally interpreted by us as showing a significantly reduced exchange constant or spin-wave velocity when considering data only below $\sim$ 100 meV.  Our new data,  Fig. \ref{dispersion}, taken with a larger incident energy extends the dynamic range of our previous results up to $\hbar \omega$ $\sim$ 200 meV.  At high-energies the momentum of the magnetic peaks agrees well with  the previous measurements on heavily underdoped and superconducting compounds and insulating YBCO$_{6+x}$ namely linear spin-waves rising from the commensurate antiferromagnetic position at $\vec{Q}$=(0.5 0.5).  The effective exchange constant at high-energies is larger and similar  to the insulating value.

	An alternate interpretation of the data in Fig. \ref{dispersion}, is that the high-energy wave vectors extrapolate back to the incommensurate wave vector at the $\hbar \omega$=0 elastic position.  This interpretation is shown by the dashed line for which the spin-wave velocity, derived from fitting all the data above $\hbar \omega$ = 40 meV, is given by $\hbar c$=800 $\pm$ 50 meV \AA.  This value is about 1.3 times larger than the value of the velocity of $\sim$ 600 meV \AA\ derived from the insulator.  This velocity would correspond to an exchange constant of $\sim$ 150 meV.  While this number seems quite large compared with the parent YBCO$_{6.0}$ and La$_{2}$CuO$_{4}$ compounds, it is similar to exchange constants derived in Nd$_{2}$CuO$_{4}$ and SrCuO$_{2}$.~\cite{Bourges97:79,Zaliznyak04:93}  This model of the spin-excitations provides a good description of all of the data above $\sim$ 50 meV.    
	
	The extrapolation back to the incommensurate position is done to relate to the stripe model where the antiferromagnetic regions are separated from hole-rich regions.~\cite{Kivelson03:75}  It is then expected that cones of scattering should exist at low-energies below the commensurate 33 meV resonance. However it has been shown in Ref. \onlinecite{Yao06:97} that for weakly coupled stripes the intensity of the inner branch is much larger than the outer branch.  Such a stripe model could then provide an explanation for the inward type of dispersion observed here and in other cuprate materials.

\begin{figure}[t]
\includegraphics[width=8.3cm] {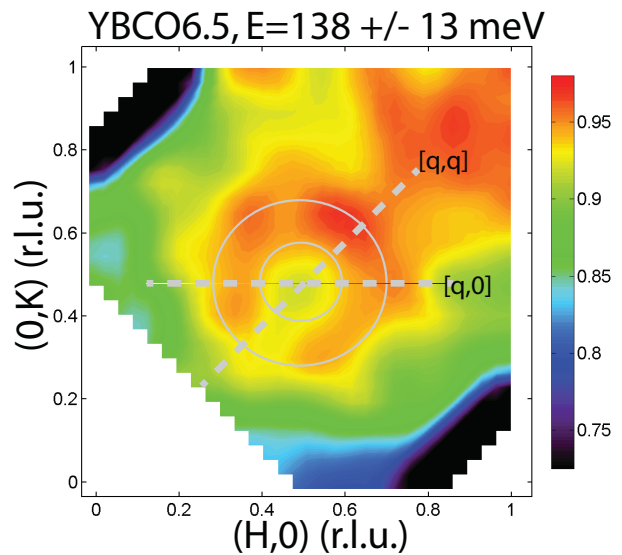}
\caption{\label{circle_colour} (Color online) A smoothed constant energy slice through the E$_{i}$=400 meV data at $\hbar \omega$=138 $\pm$ 13 meV in the superconducting phase at 10 K.  The dashed lines indicate the direction of one-dimensional cuts discussed in the text.  The increased in intensity at larger momentum transfers (near $\vec{Q}$=(1,1)) is the result of multiple phonon scattering present at higher energies.}
\end{figure}

\subsection{Pattern of spin fluctuations in momentum - isotropic planar rings above 33 meV}

	We now discuss the detailed momentum dependence of the magnetic excitations at a fixed energy transfer.  For two-dimensional spin waves, we would expect rings of scattering that are approximately isotropic in the $a^{*}-b^{*}$ plane for a fixed constant energy slice.  The shape of the scattering in momentum would change only at very high energy transfers near the top of the band where the spin-waves approach the zone boundary.  Fig. \ref{dispersion} illustrates that we are far from this limit and well below the top of the expected one-magnon band. ~\cite{Coldea01:86,Hayden96:54} 

\begin{figure}[t]
\includegraphics[width=8.3cm] {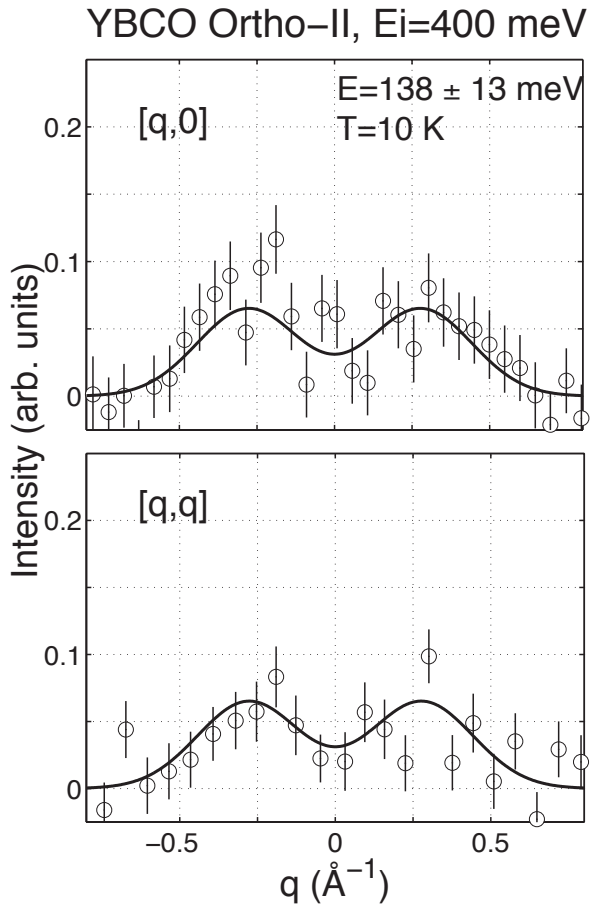}
\caption{\label{circle_cuts} One-dimensional cuts through the correlated magnetic scattering displayed in the constant energy slice in Fig. \ref{circle_colour}.  The data has been corrected for the sloping background described in the text.  The upper panel is a cut along the [1 0] direction and the lower panel is a cut along the [1 1] direction.  Both plots are displayed in terms of absolute momentum transfer (\AA$^{-1}$) and integrated over $\pm$ 0.08 \AA$^{-1}$ and $\pm$ 0.12 \AA$^{-1}$ for [1 0] and [1 1] cuts respectively.}
\end{figure}

	In contrast to spin-wave theory that gives isotropic circles of scattering it has been suggested that the scattering could form a square in the case of ladder type excitations or when scattering originates from Fermi surface nesting.~\cite{Norman00:61,Norman01:63,Si93:47,Liu95:75,Kao00:61,Tch01:63}  The interpretation of the scattering in terms of squares has been proposed by several groups most notably for La$_{2-x}$Ba$_{x}$CuO$_{4}$ and also YBCO$_{6.6}$. We shall now consider which of these models describes our new data.

	A constant energy slice through the magnetic scattering $\hbar \omega$ = 138 $\pm$ 13 meV is displayed in Fig. \ref{circle_colour}.  The data shows a ring of intensity centered at the antiferromagnetic position of $\vec{Q}$=(0.5, 0.5) plus a significant background which increases steadily with momentum transfer $\vec{Q}$ and  is the result of multi-phonon scattering.  While causing a sloping background, the magnetic scattering above background is symmetric to within experimental around the $\vec{Q}$=(0.5, 0.5) position consistent with a constant ring of scattering.

	To illustrate the isotropy, we display in Fig. \ref{circle_cuts} one-dimensional cuts through the ring of scattering of Fig. \ref{circle_colour} in two directions at 45$^{\circ}$ apart.  The upper panel is a cut through the [1 0] direction while the lower panel is a cut through the [1 1] direction plotted as a function of absolute momentum transfer (\AA$^{-1}$) relative to the antiferromagnetic position.  The sloping background has been subtracted to allow a direct comparison.  As illustrated in Fig. \ref{1dcuts}, a constant plus a slope provides an excellent description of the background over the entire energy range of these experiments.  The solid lines are the sum of two Gaussians symmetrically displaced from the antiferromagnetic point.  It can be seen that the data is well described by a model where the scattering is from isotropic spin-waves.  Moreover, the spin momentum pattern in YBCO$_{6.5}$ is inconsistent with a pattern of squares.  If it were so for YBCO$_{6.5}$, the peaks that occur at $q_{0}$=0.28 $\AA^{-1}$ along the [1 0] direction would result, for the square pattern, in a peak at 0.40 $\AA^{-1}$  along the [1 1] direction.  This is not supported by our data and we conclude that the momentum pattern at high-energies is circular as expected for an isotropic cone of spin waves intersecting a plane of constant energy.  The same conclusion was obtained earlier based on more limited data around $\hbar \omega$=78 $\pm$ 8 meV for YBCO$_{6.5}$ and also for the magnetic excitations of heavily underdoped YBCO$_{6.35}$.

	This conclusion differs from that of several high-energy studies on the monolayer and bilayer hole-doped cuprates.  For much more heavily doped YBCO$_{6+x}$, Refs. \onlinecite{Hayden04:429} and \onlinecite{Hinkov07:3} have reported a 45$^{\circ}$ rotation of the incommensurability above the resonance energy with respect to the low-energy incommensurate peaks for YBCO.  The low-energy incommensurability below the resonance along the $a$-direction in both samples is similar and corresponds to $\delta$=0.1 while the   incommensurability for our sample is $\delta$=0.06 r.l.u.  Since the incommensurability increases with doping, this shows that our samples are more underdoped than those materials where strong anisotropy has been observed.  It is therefore possible that with increasing doping, the nature of the high-energy magnetic scattering changes.  This is predicted by both Fermi surface nesting arguments and also stripe models (Ref. \onlinecite{Yao06:73,Yao08:77}) which show high-energy magnetic scattering to be very sensitive to the parameters of the theories. 

\begin{figure}[t]
\includegraphics[width=8.5cm] {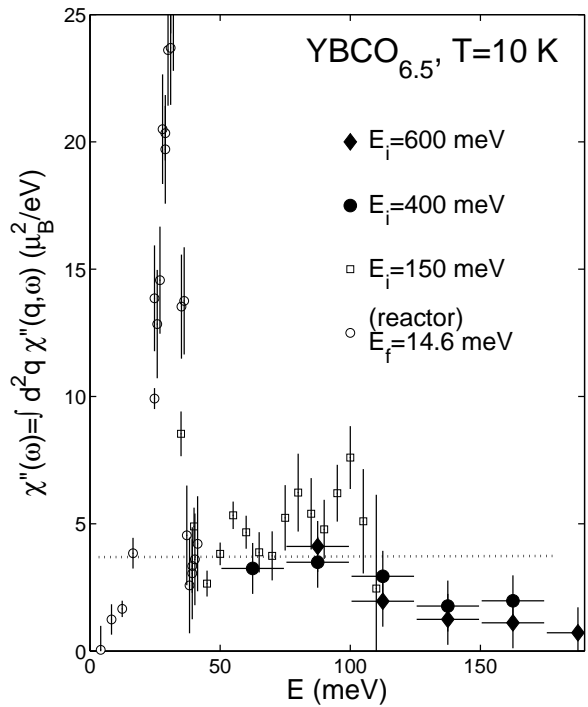}
\caption{\label{integrate} The momentum-integrated intensity (per two Cu$^{2+}$ spins) is plotted as a function of energy transfer from measurements using both reactor and spallation neutron sources.  The reactor and the E$_{i}$=150 meV data was previously published.~\cite{Stock04:69,Stock05:71}  The current data taken with E$_{i}$=400 meV and 600 meV is illustated by filled circles and diamonds respectively.  The data is integrated over the in-plane wave vectors along the $a^{*}$ and $b^{*}$ directions.}
\end{figure}

\subsection{Intensity - Suppression of spectral weight above 125 meV}

	The spectrum of the local susceptibility for YBCO$_{6.5}$ ($\chi''(\omega)=\int d^{2}q \chi''(q,\omega)$)  from the magnetic scattering integrated over the two-dimensional wave vector is displayed in Fig. \ref{integrate}.   It summarizes all data we have taken at reactor (open circles) and at spallation sources (for E$_{i}$=150 meV, 400 meV, and 600 meV).  The data was converted to absolute units through the use of a vanadium standard, as discussed previously, and is the absolute scattering cross section per formula unit of YBCO$_{6.5}$.  This is the scattering cross section for both bilayers and therefore for 2 Cu$^{2+}$ spins.  We have not integrated over the third direction in wave vector because the scattering is assumed to be two-dimensional and the intensity forms a rod along the $c^{*}$ direction while well defined in the $a^{*}$-$b^{*}$ plane.  This assumption is valid at energies above $\sim$ 40 meV as demonstrated in Ref. \onlinecite{Stock05:71}.  The high-energy (E$_{i}$=400 meV and 600 meV) values were obtained by numerically integrating constant energy cuts similar to those represented in Fig. \ref{1dcuts}. 

	Spin-wave theory for a two-dimensional antiferromagnet gives a constant momentum integrated intensity at least in the region where the energy of the excitations varies linearly with wavevector.  The proportionality constant requires the knowledge of the renormalization constant $Z$ which has been measured and theoretically found to be $\sim$ 0.5 for the case of the Cu$^{2+}$ spins in the cuprates.~\cite{Singh89:39}  This has been experimentally confirmed using neutron scattering for the insulating, parent antiferromagnets YBCO$_{6.15}$ and La$_{2}$CuO$_{4}$.~\cite{Hayden96:54,Coldea01:86} The expected constant value of the momentum-integrated cross section (as shown in Fig. \ref{integrate}) is $\chi''$=3.5 $\mu_{B}^{2}/eV$ for two Cu$^{2+}$ spins in the unit cell as shown by the dotted line in Fig. \ref{integrate}.

	The momentum-integrated intensity at low-energies, below the 33 meV resonance peak, has been discussed extensively.~\cite{Stock05:71, Stock04:69}  While the resonance peak is strongly suppressed on entering the normal phase above 59 K, we have previously observed that there is a weak remanent peak present at this energy in the normal state at 85 K (T$_{c}$=59 K).  Given the uncertainty in the origin of the resonance peak and is relation to superconductivity, we focus our attention on the magnetic spectrum above the resonance energy.  Data from MAPS using an incident energy of 150 meV showed that the momentum-integrated intensity is roughly constant between the resonance energy and below $\sim$ 100 meV, and with the magnitude in agreement with spin wave theory.  However, at high energy transfers above $\sim$ 100 meV, we now find that there is a marked and systematic decrease of the integrated intensity as shown by the data with E$_{i}$=400 meV and 600 meV in Fig. \ref{integrate}.  This is in direct contrast with linear spin-wave theory and the antiferromagnetic Mott insulator where the momentum-integrated intensity has been shown to be at least constant and then to peak at the top of the one-magnon band.  

	This systematic suppression of fast spin fluctuations mirrors and reinforces our findings for the heavily underdoped YBCO$_{6.35}$ (T$_{c}$=18 K).  There, we found that the high-energy fluctuations were damped in energy, lost significant spectral weight in comparison to the insulator, and were heavily renormalized to lower energies at the zone boundary.~\cite{Stock07:75}  This result was interpreted in terms of the spin-waves interacting with particle-hole pairs at high-energies above the pseudo-gap energy extracted from various transport measurements.~\cite{Sutherland03:67}  For YBCO$_{6.35}$ (T$_{c}$=18 K), suppression of spin fluctuations occurred at around $\hbar \omega \sim $ 200 meV, while for more heavily doped YBCO$_{6.5}$ (T$_{c}$=59 K) suppression occurs at a much lower energy of $\sim$ 100 meV.   Because the upper energy scale decreases as the doping increases we propose that the upper energy where the intensity suppression occurs follows the pseudogap.

	The loss of intensity at large energy transfers when the magnetic excitations are displaced significantly from the magnetic zone center has been observed in several metallic ferromagnets (such as Ni, Fe, and MnSi).~\cite{Paul88:38,Ishikawa77:16}  In these systems the spin excitations become broad when both the momentum and energy transfer match those of the particle-hole pairs.  A similar idea was also theoretically postulated to occur in lightly hole doped cuprates.~\cite{Onufrieva94:50}  In contrast to the observation in magnetic metals, we observe a disappearance of the excitations and not a redistribution of spectral weight.  This result is particularly surprising as it is in violation of the total moment sum rule which states that the total integrated spectral weight must be a constant.  It is possible that the spectral weight is very broad and shifted in energy resulting in it not being distinguishable from background scattering.   

\begin{figure}[t]
\includegraphics[width=8.3cm] {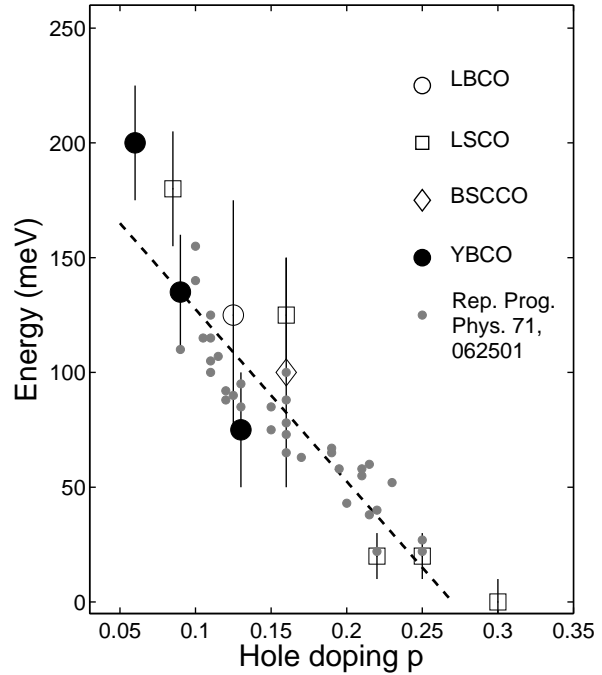}
\caption{\label{summary_cuprates} A summary of the energy value at which the momentum-integrated susceptibility ($\chi''(\omega)$) falls to half the value predicted by spin-wave theory (taken to be 1.75 $\mu_{B}^{2}/eV$ $\times$ 1/2 = 0.88 $\mu_{B}^{2}/eV$ per copper spin) for a variety of hole doped cuprate materials.  The light grey symbols are taken taken from a variety of techniques and summarized in Ref. \onlinecite{Hufner08:71}.}
\end{figure}

	To confirm this point, and to directly compare our data with other cuprates, we have carried out an extensive literature search of the neutron scattering experiments on hole doped cuprates where absolute normalization has been carried out and the energy value at which the high-energy integrated intensity (as plotted in Fig. \ref{integrate}) falls to $1/2$ of the value predicted from linear spin-wave theory.  We have taken this half height value to be 1.75 $\mu_{B}^{2}/eV$ per two Cu$^{2+}$ spins (or 0.88 $\mu_{B}^{2}/eV$ per Cu$^{2+}$ spin). 

	Where there are inconsistencies for a particular hole concentration, we have taken the most recently published data for a given hole doped concentration.  Most of the data has been taken with the MAPS spectrometer at ISIS, and so the data is at least self-consistent so far as the calibration to absolute units.  We have previously compared (Ref. \onlinecite{Stock05:71}) the absolute normalization for data taken on MAPS (using the vanadium standard) to an internal calibration based on an acoustic phonon measured using a thermal triple-axis spectrometer and have found agreement within $\pm$ 20 \%.  The only exception is the data taken on nearly optimally doped YBCO$_{6.85}$ which was taken with a triple-axis crystal spectrometer and no absolute calibration was performed.  Here we have taken the energy to be 75 $\pm$ 25 meV based on the constant energy scans presented in the publication.~\cite{Pailhes04:93}   

	The data are plotted as a function of hole doping in Fig. \ref{summary_cuprates} for hole doped cuprates YBa$_{2}$Cu$_{3}$O$_{6+x}$ (Refs. \onlinecite{Pailhes04:93} and \onlinecite{Stock07:75}), La$_{2-x}$Sr$_{x}$CuO$_{4}$ (Refs. \onlinecite{Lipscombe07:99,Waki07:98,Vignolle07:3,Lipscombe09:102,Christensen04:93}),  La$_{2-x}$Ba$_{x}$CuO$_{4}$ (Refs. \onlinecite{Tranquada04:429} and \onlinecite{Xu07:76}), and Bi$_{2}$Sr$_{2}$CaCu$_{2}$O$_{8+\delta}$ (Ref. \onlinecite{Xu09:5}).  The value of the hole doping $p$ was obtained from the formula proposed in Ref. \onlinecite{Tallon95:51}.  The light grey points are pseudogap values taken from a variety of techniques on a series of cuprates outlined in Ref. \onlinecite{Hufner08:71} and the references therein.  The dotted line shows the curve proposed in Ref. \onlinecite{Hufner08:71} ($E_{pg}=E_{\circ}(0.27-x)/0.22$) to describe the doping dependence of the pseudogap energy value.   The trend of the magnetic data we present in Fig. \ref{summary_cuprates} is in good agreement with this pseudogap formula with a value of E$_{\circ}$=165 $\pm$ 10 meV for the maximum pseudogap value in comparison to 152 $\pm$ 8 meV obtained in Ref. \onlinecite{Hufner08:71} to describe a series of transport, optical, and tunneling data.  The agreement between neutron and electronic probes is quite good especially given our pragmatic definition of the pseudogap as the value being the energy where the neutron integrated intensity falls to $1/2$ the value predicted by linear spin-wave theory.  Our values for the trend of the pseudo-gap energy are consistent with those of Ref. \onlinecite{Hufner08:71} and enable us to conclude that the decrease in the high energy intensity suppression tracks and results from the decay of spin fluctuations to particle-hole pairs above the pseudogap energy.   It is interesting to note that while the lower energy resonance only scales with doping and concentration over a limited range near optimal doping, the high-energy scale tracks universally across all hole concentrations and cuprate concentrations.  It is also consistent with probes directly sensitive to the electronic properties.

	Our experimental studies of the spin fluctuations therefore indicate the presence of two distinct energy scales in the cuprate superconductors.  The first energy scale is associated with the resonance peak and scales with superconductivity over a narrow range of doping and is not a universal energy scale across different cuprates and hole concentrations, as discussed in the introduction.  Our high-energy work points to the presence of an additional energy scale which tracks the pseudogap energy and is consistent amongst different cuprate materials and hole concentrations.  There is considerable experimental evidence for such a scenario from ARPES (Ref. \onlinecite{Tanaka06:314}), Raman (Ref. \onlinecite{Guyard08:101}), and tunnelling measurements (Ref. \onlinecite{Pasupathy08:320,Boyer07:3}) and possible physical mechanisms have been discussed in Ref. \onlinecite{Chien09:79,Norman07:76}.  The fact that we observe anomalous spin dynamics near the magnetic zone boundary may imply that this second energy scale is associated with electronic effects near the anti nodal positions.  ARPES measurements have found evidence for such an energy scale which does not scale with the superconducting T$_{c}$ and decreases with increasing doping (Ref. \onlinecite{Damascelli03:75}), consistent with the present experimental results from neutron scattering.   The energy scale was observed as a broad peak (termed the hump) along the ($\pi$,0) direction which remains above the superconducting transition temperature.~\cite{Param04:79}  Therefore, we find the upper spin energy scale, observed here with neutrons, tracks the doping charge hump energy gap measured using ARPES.

\subsection{High energy mode - Detection, experimental concerns,  and possible origins}

\begin{figure}[t]
\includegraphics[width=8.7cm] {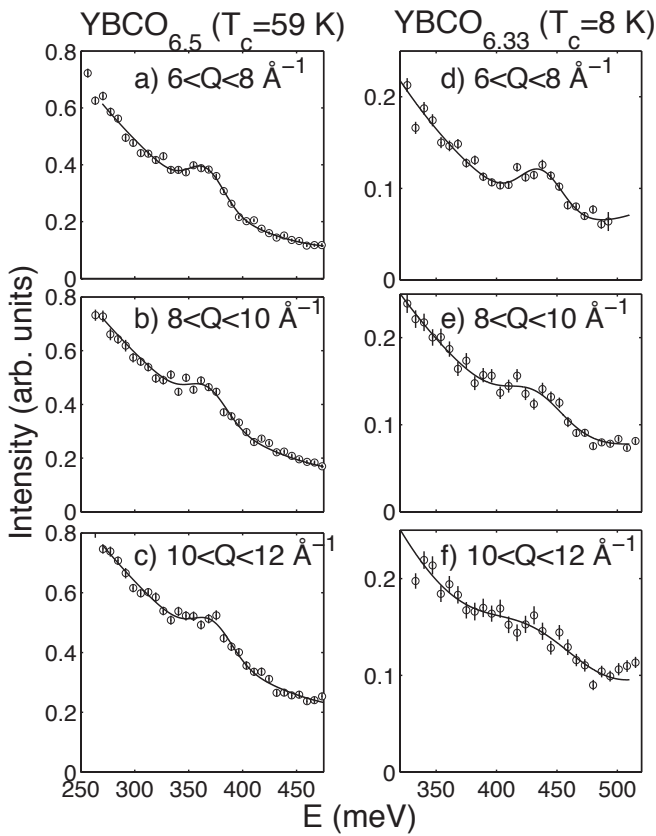}
\caption{\label{highe_cuts} Constant $Q$ cuts taken on the MARI direct geometry spectrometer with an incident energy of 750 meV and with the relaxed chopper spinning at 600 Hz.  Panels $a)$-$c)$ are for YBCO$_{6.5}$ (T$_{c}$=59 K) and panels $d)$-$f)$ for YBCO$_{6.33}$ (T$_{c}$=8 K). The solid line is a fit to a gaussian function on a sloping background.}
\end{figure}

	So far we have been concerned with the properties of the magnetic excitations within the band of the lowest energy excitation (one-magnon band). In this section, we examine the possibility of a magnetic  response well above the upper energies associated with the one-magnon band at $\hbar \omega \sim$ 250-300 meV.  While this is a relatively unreported region, there are several noteworthy x-ray and optical studies which reflect the need to explore this energy range with neutrons.   Measurements of the absorption coefficient using mid-infrared spectroscopy have reported a strong peak at an energy of about 425 meV for insulating La$_{2}$CuO$_{4}$ and at a slightly lower energy of $\sim$ 375 meV for Sr$_{2}$CuO$_{2}$Cl$_{2}$.~\cite{Perkins98:58}  Relatively recently a peak was observed in La$_{2}$CuO$_{4}$ using resonant inelastic x-ray scattering at an energy transfer near 500 meV.~\cite{Hill08:100,Ellis10:81}  While several explanations were proposed, the origin of this excitation is unclear.  Since neutron scattering has both a well defined cross section and selection rules, it may help in identifying the origin of any magnetic peaks at these high energies.

\begin{figure}[t]
\includegraphics[width=8.7cm] {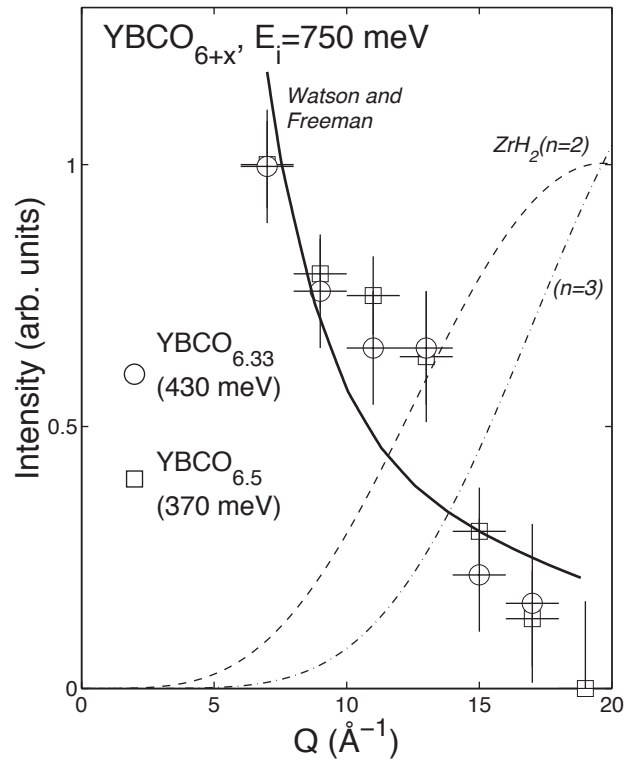}
\caption{\label{figure_ff} The intensity of the high-energy peak measured on MARI for YBCO$_{6.33}$ (T$_{c}$=8 K, open circles) and YBCO$_{6.5}$ (T$_{c}$=59 K, open squares) as a function of momentum transfer.  The form factor from Ref. \onlinecite{Watson61:14} is shown as a solid line.  For comparison, the intensity as a function momentum transfer of the excited states of ZrH$_{2}$ are also plotted to illustrate that the intensity is inconsistent with a hydrogen related molecular excitation.}
\end{figure}

\begin{figure}[t]
\includegraphics[width=8.7cm] {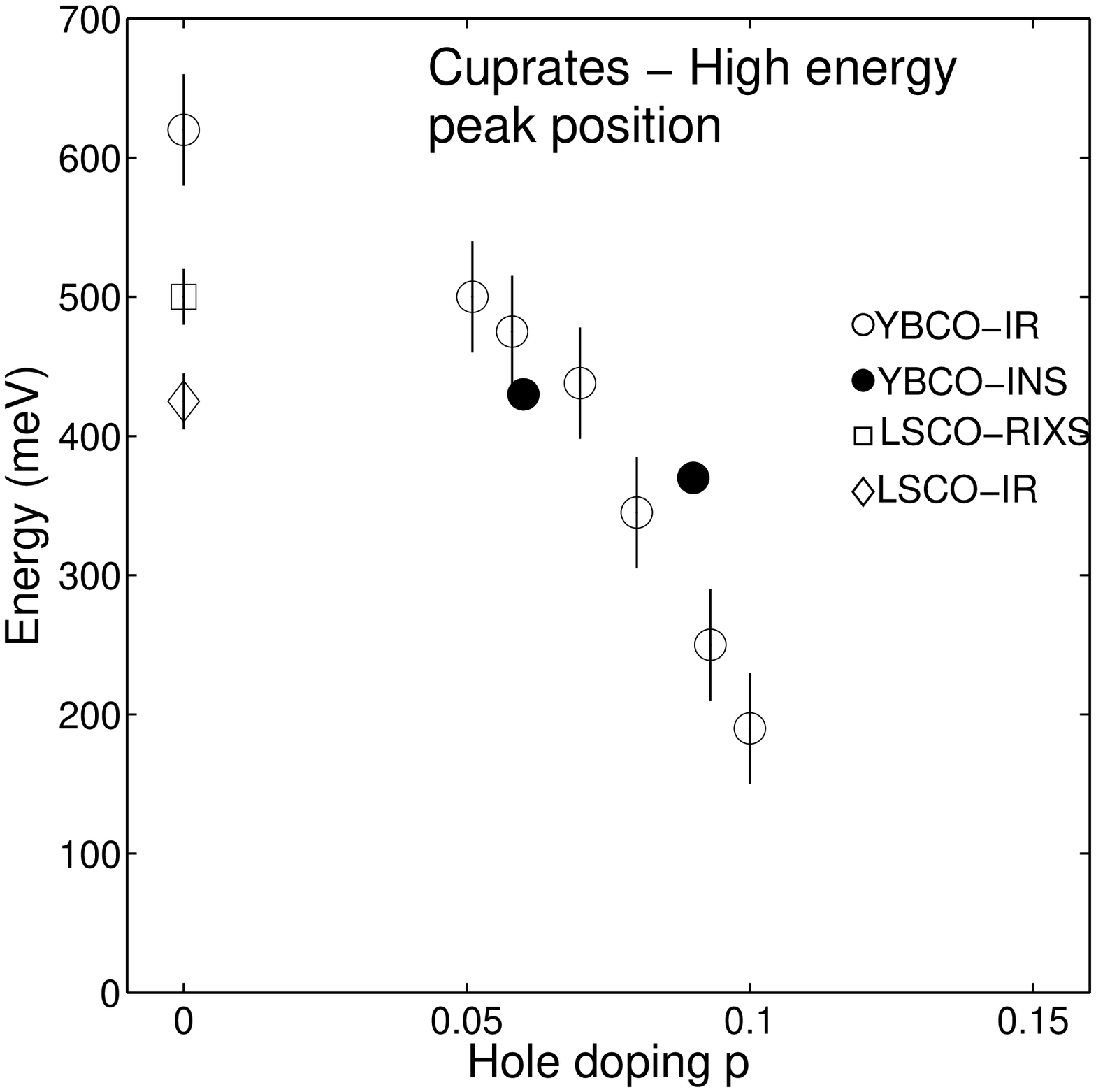}
\caption{\label{summary_RIXS} The energy of the high energy mode is plotted as a function of hole doping for both LSCO and YBCO systems.  The YBCO inelastic neutron scattering (INS) data is described here, the YBCO mid-infrared was taken from Ref. \onlinecite{Lee72:05}, the LSCO resonant inelastic x-ray scattering (RIXS) is from Ref. \onlinecite{Hill08:100}, and the LSCO mid-infrared results are from Ref. \onlinecite{Perkins98:58}.}
\end{figure}

	A search for these high-energy excitations was initially made using the MAPS spectrometer on YBCO$_{6.5}$ and evidence was found for a peak at an energy transfer near about 370 meV.  Due to the limited detector range on MAPS, we continued these measurements on MARI on samples of YBCO$_{6.5}$ (T$_{c}$=59 K) and YBCO$_{6.33}$ (T$_{c}$=8 K).  The results from the MARI experiment with an incident energy of 750 meV are plotted in Fig. \ref{highe_cuts} for both YBCO$_{6.5}$ and YBCO$_{6.33}$.  Both experiments on MARI were performed at room temperature without using a closed cycle refrigerator.  

	A peak is confirmed for YBCO$_{6.5}$ (T$_{c}$=59 K) at 370 meV and is visible at low momentum transfer. As the momentum transfer increases the intensity decreases.  A similar result is found for YBCO$_{6.33}$ (T$_{c}$=8 K), but with the peak shifted to a higher energy transfer of 430 meV and their energies plotted in Fig. \ref{summary_RIXS}.  The solid lines in Fig. \ref{highe_cuts} are fits to a Gaussian on a sloping background.

	One reason for concern is that the peak energy is comparable to vibrational frequencies from water or metallic hydrides that may be present in the sample as impurities.~\cite{Bennington93:181,Li92:4,Bennington91:172,Bennington90:151}  Unfortunately, YBCO$_{6+x}$ may absorb small amounts of water.  Given the large neutron cross section for hydrogen, even minute amounts of hydrogen could provide a significant cross-section which could be erroneously interpreted  as magnetic.  The problem is even more complicated because the molecular modes involving hydrogen do not scatter at high-energies beyond a scattering angle of $2\theta$=90$^{\circ}$ because at large energy transfers the impulse approximation ensures that each scattering site behaves as an independent source.  This approximation has been reviewed and studied in detail in Ref. \onlinecite{Stock10:81}.  As a result the standard rule that vibrational scattering grows continuously with momentum transfer while magnetic scattering does not, must be treated with caution.

	We address this point in Fig. \ref{figure_ff} where the intensity as a function of momentum transfer was extracted from the cuts shown in Fig. \ref{highe_cuts} and the two data sets for YBCO$_{6.33}$ and YBCO$_{6.5}$ were normalized to agree for the lowest momentum transfer studied.   The solid line is the calculated Cu$^{2+}$ form factor from Ref. \onlinecite{Watson61:14} and this agrees well with the data.  

	The dashed lines are the results of a calculation using harmonic theory for the momentum dependence of a molecular excitation involving hydrogen measured for ZrH$_{2}$.~\cite{Li92:4}  We have chosen ZrH$_{2}$ as the experimental data have been extremely well characterized and modelled and also lower energy excitations ($\sim$ 150 meV) have been found to be quite similar in nature to those measured in YH$_{x}$.  The $n=2$ and $n=3$ excitations plotted in Fig. \ref{figure_ff} correspond to energies of 294 meV and 441 meV and therefore represent examples of the scattering from the molecular excitations in this energy range.~\cite{Ikeda90:2}  We can see that whereas the momentum dependence of the molecular excitations is peaked at a finite momentum transfer, for momentum transfers below $\sim$ 15 \AA$^{-1}$, the intensity is inconsistent with our experimental data.

	Based on these results, there is considerable evidence against the excitations we observe resulting from molecular modes involving hydrogen.   Firstly, the energy of the excitation changes by $\sim$ 50 meV with a very small change in the oxygen concentration.  Secondly, the momentum dependence is inconsistent with the previous measurements on metal hydrides at similar energies.  Thirdly, if the excitation did originate from a molecular excitation involving hydrogen, then we would expect lower energy excitations which have stronger structure factors than the excitation at $\sim$ 400 meV.  A search revealed no such peaks.  We therefore suggest that these results are magnetic and are associated with scattering from the Cu$^{2+}$ ions. 

	Despite these facts, we do note that a recent study on MARI of La$_{2}$CuO$_{4}$ did observe a hydrogen related mode at $\sim$ 450 meV which did not change energy with doping.~\cite{Kim_unpub}  While conclusive proof for the existence of an excitation will only come from corroborating evidence from other techniques, it is interesting to speculate as to the possible magnetic and electronic origins of this excitation. The energy of the peak approximately agrees with a variety of other techniques.  The measured value of 430 meV for heavily underdoped YBCO$_{6.33}$ agrees quite well with the optical studies on the parent La$_{2}$CuO$_{4}$ compound in Ref. \onlinecite{Perkins98:58}.  The energy is also in approximate agreement with a variety of resonant inelastic x-ray studies that report a peak at 500 meV energy transfer.~\cite{Hill08:100,Ellis10:81} The substantial decrease in the energy with hole doping has also been reported in Ref. \onlinecite{Braicovich08:xx} where the energy of the high-energy excitation was observed to decrease from 500 meV to below 300 meV as the doping varied from the parent undoped compound to optimal concentrations.  We summarize the energy positions measured by resonant x-rays, mid-infrared, and our neutron measurements for both LSCO and YBCO in Fig. \ref{summary_RIXS}.  For YBCO$_{6+x}$, the hole doping has been extracted from the value of T$_{c}$ using the formula described in Ref. \onlinecite{Tallon95:51}.  For consistency, we have omitted the data where only a Neel temperature exists except for the trivial case of $p=0$.  The trend with doping in Fig. \ref{summary_RIXS} suggests that the peaks measured with neutrons and optical techniques have a common origin and also that there is a consistent softening with hole doping.  We note that mid infrared experiments observed the peak to be very broad in energy while we observe a relatively sharp excitation in energy.~\cite{Lee72:05}

	Possible magnetic mechanisms for this peak include a crystal field ($dd$) type transition that might be expected from an orbital transition, or from a bi-magnon excitation suggested in papers on resonant inelastic x-ray scattering.  The bi-magnon excitation is not the traditional two-magnon process investigated using neutron scattering.  It was shown in Ref. \onlinecite{Lorenzana95:74} that a finite magnon-magnon interaction can result in a resonant excitation at high-energies.   It is interesting to note that the peak is predicted to occur at $\sim$ 2.7 $J$ which is $\sim$ 370 meV, close to the energy observed here for YBCO$_{6.5}$. 
	
	An orbital  magnetic transition is also a possible origin for the high energy peak.  We show in the Appendix that the local electrostatic field splits copper states into 5 orbital singlets coupled to the spin.  The result suggests that with the crystal field parameters arising from the local oxygen structure, the first excited orbital state could have the observed energy. Clearly, further studies are required to determine the underlying origin of this excitation and its relevance to the physics of the cuprates.  

\section{Conclusions and summary}

	We have investigated the high-energy spin response in the underdoped Ortho-II YBCO$_{6.5}$ superconductor.  Our most important finding is the presence of a second high-energy scale in the spin fluctuations which is characterized by a suppression of spectral weight.  A comparison with techniques which directly probe the electronic properties illustrates that this second energy scale is associated with the pseudogap.   Within the error of the measurements, this second energy scale seems to scale universally across all cuprate compositions and dopings.  Based on these findings, we find direct evidence for the pseudogap in the spin fluctuations and evidence for the presence of two distinct energy scales in the cuprate superconductors.

	We have also searched for excitations above the top of the one-magnon band measured to be 300-350 meV in the parent La$_{2}$CuO$_{4}$ compound.  We have observed a well defined excitation at 370 meV in YBCO$_{6.5}$ (T$_{c}$=59 K) and a similar excitation at 430 meV in YBCO$_{6.33}$ (T$_{c}$=8 K).  The intensity as a function of momentum transfer is consistent with a magnetic excitation and not in agreement with expectations of molecular excitations involving hydrogen known to exist in a similar energy range.  This was demonstrated through a comparison to ZrH$_{2}$, a typical and well characterized metal hydride.  Excitations of similar energy have been observed in the mid-infrared region and also using resonant inelastic x-rays.  While possibilities include an orbital $dd$ transition and bimagnon type fluctuations, more work is required to understand the underlying mechanism and true nature of these excitations. 

\begin{acknowledgements}

We are grateful for the assistance of K. Allen and A. Orsulik for expert technical support to Y.-J. Kim and R. Coldea for helpful discussions.  RAC was partially supported by a Leverhulme grant.

\end{acknowledgements}

\section{Appendix: Calculation of spin and orbital excitations in YBCO$_{6+x}$}

	We have calculated the energies of high-energy transitions of the copper spin and orbital moments with the full symmetry of the local electric field environment and with spin-orbit coupling.  We find that the magnetic dipole $dd$  transitions occur close to the energies of peaks observed in neutron scattering experiments.

	It is commonly assumed that the magnetic properties of the Cu$^{2+}$ ion in cuprate superconductors may be described by a spin-one-half doublet in which the orbital magnetic moment is quenched.  This assumption may give a good approximation for low-energy spin waves, being transitions between the two members of the spin doublet, but it fails to describe magnetic properties in which the orbital moment is active. The 9 electrons (or one hole) in the $3d$ shell of Cu$^{2+}$ form a Hund's rule state with spin $S=1/2$ and orbital angular momentum L = 2 coupled by the spin-orbit interaction

\begin{eqnarray}
\label{spin_orbit}
H_{SO}= \lambda \vec{L} \cdot \vec{S}
\end{eqnarray}

\noindent where $\lambda$=-102.9 meV for free ions.~\cite{Abragam70:book} 

The five L = 2 states, if subject to a local electric field having the cubic symmetry of an octahedron of neighbors, would be split into a ground doublet and excited triplet.  A splitting pattern of similar symmetry occurs if the interaction is regarded as arising from the hybridization of the Cu$^{2+}$ 3d states with the 2p-states of the O$^{2-}$ neighbors.  In the lower symmetry that prevails in YBCO one of the octahedral vertex neighbours in the $z$ direction is missing and the apical oxygen has only a minor effect on the Cu$^{2+}$ orbital ground state.  The remaining four oxygen neighbors lie close to but are buckled out of the $xy$ copper plane.  This gives a strongly tetragonal field that splits the lowest cubic doublet by 338 meV into the two orbital singlet states  the lowest being $x^{2}-y^{2}=(\mid 2 \rangle +\mid -2 \rangle)/\sqrt{2}$ as well as splitting levels of the upper triplet (see Table I, left hand column).

We have placed the neighbors at the low-symmetry positions of the known crystal structure.~\cite{Jorgensen90:41}  This gives the spectrum of excited states without the common assumption that a symmetric octahedral field dominates.  The relative splitting pattern is expected to be reasonable while the overall magnitude of the splitting is expected to be less reliable.  We regard the calculations as the lowest-order estimate of the energies of the excited orbital states.    

\begin{figure}[t]
\includegraphics[width=8.3cm] {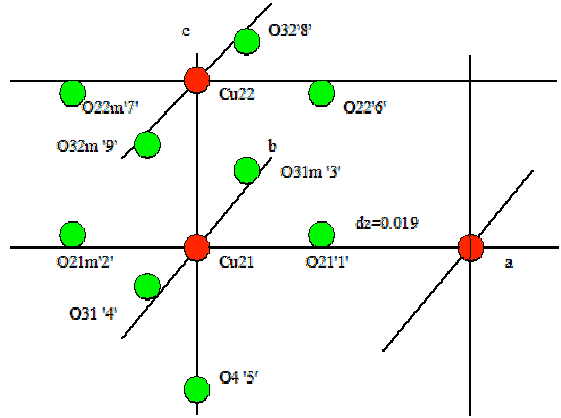}
\caption{\label{figure_cef} (Color online) Local structure about a central copper ion ($Cu21$) in YBCO$_{6+x}$.  In the buckled $a-b$ plane the neighbors of the $Cu21$ atom (red) are four nearest neighbors oxygens, $O2$ and $O3$ (green) displaced up from the plane by $dz$, one apical oxygen ($O4'5'$ green) and four oxygens ($O2$, $O3$) in the bilayer plane above displaced down by $dz$.  The $n$'th of the nine oxygen neighbors is denoted by the trailing label `$n$'.}
\end{figure}

	The crystalline field energies have been calculated for up to 18 near neighbors but focus is on the 4 nearest O$^{2-}$ ions which are found to dominate the spectral pattern.  The atomic structure around a central Cu$^{2+}$ ion, $Cu21$ at (0,0,0), in the lower bilayer is shown in Fig. \ref{figure_cef}.  The buckling of the $O21$ and the $O31$ ions in the $Cu(2)$ planes is included through their $z$ fractional displacement $dz$=0.019 as measured by Jorgensen.~\cite{Jorgensen90:41}  The lattice constants are those of YBCO$_{6.5}$.

\begin{figure}[t]
\includegraphics[width=8.5cm] {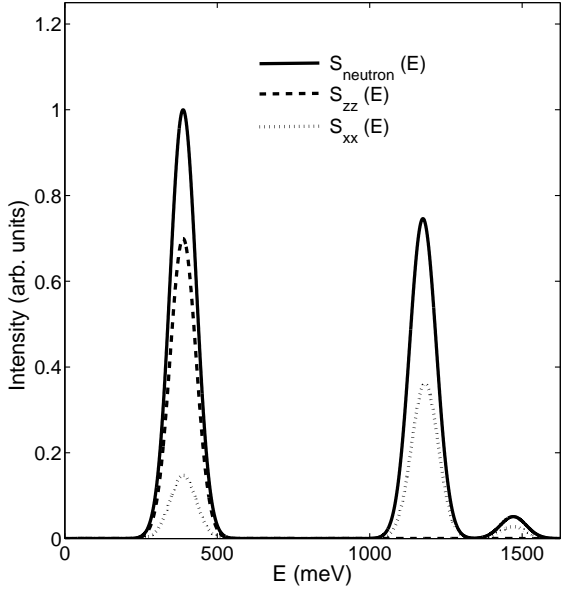}
\caption{\label{cef_1} Magnetic peaks in the dynamic susceptibility versus energy E for crystal field and spin orbit coupling in vanishing exchange field.  $S_{xx}(E)$ gives in-plane fluctuations and $S_{neutron}(E)$ gives their sum.  The longitudinal fluctuations $S_{zz}(E)$ perpendicular to the plane of the orbital moment coupled to the spin are strong.  Neutron resolution is 50 meV (full width at half height).  The energy of the lower peak at 388 meV lies close to that of the peak observed in our neutron experiments where a peak at 370 meV is observed.  The neutron resolution was set to 100 meV.}
\end{figure}

	The method is described by Hutchings.~\cite{Hutchings64:16}  The local electric field contains terms up to fourth order in the electron coordinates.  They are sums over neighbor sites $j$ of a geometric factor determined by local ionic positions and proportional to the tesseral harmonics, $Z^{np}_{m}(j)$, where $m$=2,4 and $n$=0...m. These are multiplied by an even, $p=c$, or odd, $p=s$ order Stevens operator equivalent in $\vec{L}$ expanded as a symmetrized function of the non-commuting angular momentum operators, $O^{np}_{m}(\vec{L})$.  The site symmetry emerges because particular terms in a given $Z^{np}_{m}(j)$ coefficient summed over $j$ cancel as required by symmetry.  The matrix elements of the Stevens operator equivalent yield the 5 by 5 matrix whose eigenvalues and eigenfunctions give the Cu$^{2+}$ energies and wave functions.  One example is the 4'th order potential

\begin{eqnarray}
\label{potential}
V_{4}&=&{{4\pi}\over{2.4+1}}\sum\limits_{j}^{nn} {Q_{j}\over R_{j}^{5}} \{C_{4}^{0} Z_{4}^{0}(j)O_{4}^{0}+ \nonumber \\
& & C_{4}^{1} \left(Z_{4}^{1c}(j)O_{4}^{1c}+Z_{4}^{1s}(j)O_{4}^{1s}\right)+ \nonumber \\
& & C_{4}^{2}\left( Z_{4}^{2c}(j)O_{4}^{2c}+Z_{4}^{2s}(j)O_{4}^{2s}\right)+ \nonumber \\
& & C_{4}^{3}\left(Z_{4}^{3c}O_{4}^{3c}+Z_{4}^{3s}(j)O_{4}^{3s}\right)+
\nonumber \\
& & C_{4}^{4}\left( Z_{4}^{4c}(j)O_{4}^{4c}+Z_{4}^{4s}(j)O_{4}^{4s}\right)\}
\end{eqnarray}

\begin{figure}[t]
\includegraphics[width=8.5cm] {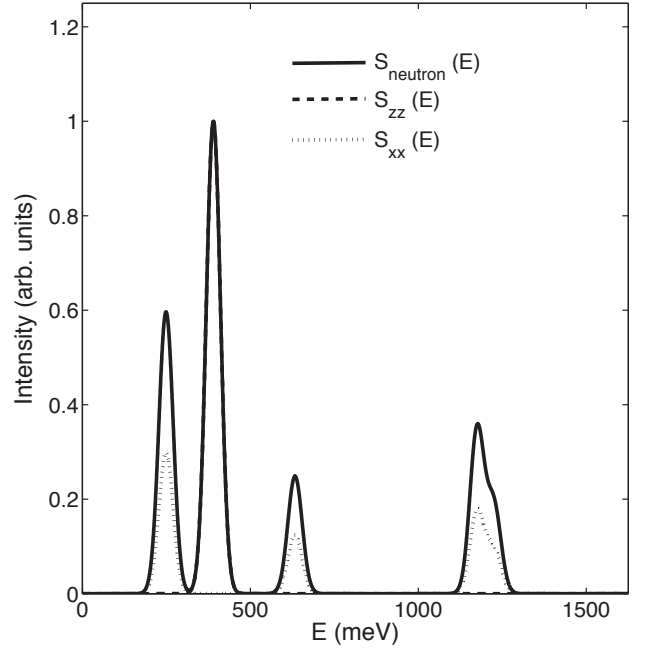}
\caption{\label{cef_2} The $dd$ magnetic transition spectrum for neutron scattering arising from crystal field, spin-orbit and exchange field of 250 meV.  The largest peak arises from fluctuations along the c-axis $S_{zz}(E)$.  $S_{xx}(E)$ gives in-plane fluctuations and $S_{neutron}(E)$ gives their sum.  The neutron resolution is 50 meV.}
\end{figure}

\noindent which enters the fourth order contribution to the Hamiltonian as $H_{4}=-e^{2}\theta_{4} \langle r^{4} \rangle a_{0}^{4} \times V_{4}$ where the Stevens factor $\theta_{4}$=-$2/63$, and $\langle r^{4} \rangle$ = 2.498 is the mean fourth power of the 3d radial wave function in Bohr radii, $a_{0}$.  The simpler second order term (not shown) was calculated with $\theta_{4}$=-$2/21$, and $\langle r^{2} \rangle$ = 1.028.  The net crystal field Hamiltonian for the YBCO$_{6.5}$ structure taking the oxygen charge to be Q$_{j}$=-2 is

\begin{eqnarray}
H_{c}=-104.9 O_{2}^{0}+3.04 O_{c}^{2c}-1.09 O_{4}^{0}+\nonumber \\
0.07 O_{4}^{2c}-14.1 O_{4}^{4c}
\end{eqnarray}

\noindent The coupling of the spin to the five orbital singlets yields a spectrum of ten $L-S$ states.  Each orbital singlet becomes a spin doublet.  The spin-orbit interaction (Eqn. \ref{spin_orbit}) serves to shift the energies of the purely orbital states.  For example, the lowest two orbital singlets, for which the orbital moment is entirely off-diagonal and equal to $L_{z}$=2, are split by 338 meV before spin-orbit coupling increases the splitting to 388 meV as shown in Table I.

\begin{table}[ht]
\label{table1}
\begin{tabular}{ l | l | l } 
\hline \hline
YBCO$_{6.5}$ & YBCO$_{6.5}$ & YBCO$_{6.5}$ nn=9 O$^{2-}$  \\ 
nn=4 O$^{2-}$ inplane & nn=4 O$^{2-}$ inplane & 4 inplane, 1 (apical), \\
 & & 4 (bilayer) \\
\hline
H$_{mfz}$=0 meV & H$_{mfz}$=250 meV & H$_{mfz}$=250 meV \\
$\lambda$= -103 meV & $\lambda$=-103 meV & $\lambda$=-103 meV \\
\hline \hline
Energy (meV) & Energy (meV) & Energy (meV) \\
\hline
0 & 0 & 0 \\
0 & 248 & 247 \\
388 & 390 & 387 \\
388 & 633 & 601 \\
1168 & 1175 & 777 \\
1168 & 1225 & 800 \\
1188 & 1314 & 815 \\
1188 & 1418 & 1023 \\
1471 & 1592 & 1133 \\
1471 & 1682 & 1159 \\
\hline
\end{tabular}
\caption{shows the weak dependence of the lowest energy levels on the number of neighboring charges (nn).  For 9 oxygen ions the results are similar to those with only the 4 nearest ions.  For the exchange-free field the lowest purely orbital level at 338 meV is moved by the spin-orbit interaction to 388 meV. Here $a$=3.83, $b$=3.86, $c$=11.78 \AA\ and the lower CuO$_{2}$ plane lies at $z=$0.360.  An exchange field in the ordered antiferromagnetic would split the spin-generate ground degenerate state by 248 meV.}
\end{table}

The crystal field and spin-orbit coupling were diagonalized simultaneously.  In Table 1, we show that the spectrum of Cu$^{2+}$ energies does not depend much on the number of neighbors.  Henceforth, we present results only for the $nn$=4 O$^{2-}$ in-plane buckled neighbors.  

The ground state wave function is then that of a pair of time-reversed states, which for spin up is 

\begin{eqnarray}
\label{wave_function2}
\mid 0,z\rangle & = &  \nonumber \\
& & 0.87\mid2,1/2\rangle-0.01\mid0,1/2\rangle+0.50\mid-2,1/2\rangle \nonumber \\
& & +0.04\mid-1,-1/2\rangle.
\end{eqnarray}

\noindent This is very closely equivalent to $\mid 0 z \rangle=0.96 \mid x^{2}-y^{2} \rangle + 0.26 \mid xy \rangle$ which reveals the persistence of the strong $x^{2}-y^{2}=(\mid 2 \rangle +\mid -2 \rangle)/\sqrt{2}$ orbital contribution.  The mixing of the only excited $xy$ state occurs because the only strong spin-orbit effect on the ground state is via a single off-diagonal matrix element $\langle xy \mid L^{z} \mid x^{2}-y^{2}\rangle=2$ coupling the orbital ground to excited states.

	The magnetic dipole spectrum is weighted by the squares of transition matrix elements between states.  The neutron spectrum is shown in Fig. \ref{cef_1}.  The peak at 388 meV is a mixture of strong longitudinal $zz$ symmetry fluctuations via $|M^{z}|^{2}$ and a spin lowering amplitude $|M^{-}|^{2}$ as listed in Table II.

\begin{table}[ht]
\label{table2}
\begin{tabular}{ l | l | l | l} 
\hline \hline
E & M$^{-}$ & M$^{+}$ & M$^{z}$ \\ \hline
0 & 0 & 0 & 2.00 \\
0 & 1.88 & 0 & 0 \\
388 & 0 & 0 & -1.74 \\
388 & 1.14 & -0.01 & 0 \\
1168 & -1.43 & 0.38 & 0 \\
1168 & 0 & 0 & -0.02 \\
1188 & 0.54 & 0.91 & 0 \\
1188 & 0 & 0 & 0.01 \\
1471 & 0 & 0 & 0 \\
1471 & -0.02 & -0.47 & 0 \\
\hline
\end{tabular}
\caption{The table lists energies and gound-state transition amplitudes for crystal field and spin-orbit states in a vanishing exchange field.}
\end{table}

The energy of 388 meV calculated for the magnetic $dd$ transition peak lies close to the observed energy of 370 meV (Fig. \ref{highe_cuts}) in YBCO$_{6.5}$.  This inelastic peak arises from direct magnetic transitions as opposed to the bi-magnon mechanism adoped by some (Ref. \onlinecite{Braicovich09:102}) to explain peaks seen in inelastic x ray scattering.  Evidence for the higher energy peaks has been obtained with both midinfrared and resonant inelastic x ray scattering therefore providing some validation of this calculation.~\cite{Falck94:49,Ellis08:77,Kim09:79}

To obtain the lower energies observed at larger doping (Fig. \ref{summary_RIXS}) we reduced the near-neighbor oxygen charge to reflect hole doping (Ref. \onlinecite{Radaelli98}).  This does lower the calculated peak energies but by less than observed in experiment, in part because $dp=0.1$ range of $p$ is small compared with the nominal $2^{-}$ charge.  We surmise that the peak energy declines with doping because of increased screening of the electric field by the growing carrier density of the electron gas.  To produce the very low peak observed with infra-red in Fig. \ref{summary_RIXS} at $\sim$ 200 meV would require an unreasonable ``effective" oxygen charge less than half nominal or a weakened hybridization.

	When magnetic order is present, the strength of the interatomic exchange is

\begin{eqnarray}
\label{inter}
H_{ex}= J \sum\limits_{i,j} \vec{S_{i}} \cdot \vec{S_{j}}
\end{eqnarray}

\noindent where $J$=125 meV is greater than the spin-orbit interaction.  For the ordered antiferromagnet a spin feels, in two dimensions, a static molecular field

\begin{eqnarray}
\label{H_exmf}
H_{mfz}= 4JS = 250 meV.
\end{eqnarray}

\noindent The ground state wave function Eq. \ref{wave_function2} for $H_{mfz}=0$ is essentially unchanged at high-field.  This is because an exchange field along $z$ couples an up spin only to the up spin component of the distant excited off-diagonal orbital state.  The result is that the ground state doublet splits symmetrically with negligible exchange mixing of excited states.

A large moment, 2.01 $\mu_{B}$, of which the orbital moment is half, is produced when an exchange field $H_{mfz}$ of 250 meV acts along the $c$ direction.  The smallness of the increase over zero field ($M^{z}$=1.997) is because the $z$ component of the orbital moment is off-diagonal between the lowest two orbital states.  The moment for a field in the plane is half as large.  The uniform susceptibility is known to be larger along the $c$ direction although NMR (Ref. \onlinecite{Walstedt92:45})gives a ratio of $\chi_{c}/\chi_{ab}$ of only 1.28 estimated from band-like anisotropy as the internal exchange and bilayer coupling of the real material serves largely to overcome the underlying single-ion easy axis.     Our results extend those of Abragam and Bleaney to relatively large spin orbit coupling, $\lambda/\Delta_{0}$ = -0.30, where the $xy$ state at $\Delta_{0}$ = 338 meV in the large tetragonal field of YBCO now lies below the $3z^{2}-r^{2}$ state of their perturbation theory.  The susceptibility normal to the plane is again ~twice that within the planes with similar but reduced g-values ($g_{z}$ = 4 ).  

When the planar spin correlation length is long enough, a given spin may feel some effect of a root mean square local field allowing spin waves to propagate, as we have found not only for YBCO$_{6.5}$ where low-energy incommensurate fluctuations extend over $\xi \sim 20$ \AA (Ref. \onlinecite{Stock04:69}), but even more so for YBCO$_{6.35}$ and YBCO$_{6.33}$ where the elastic correlation lengths reach 42 \AA (Ref. \onlinecite{Stock06:73,Stock08:77}) and 76 \AA (Ref. \onlinecite{Yamani07:430}) respectively.  This local effect we ignore for the highest energy transitions.

\begin{table}[ht]
\label{table3}
\begin{tabular}{ l | l | l | l} 
\hline \hline
E & M$^{-}$ & M$^{+}$ & M$^{z}$ \\ \hline
0 & 0 & 0 & 2.01 \\
249 & 1.89 & 0 & 0 \\
390 & 0 & 0 & -1.73 \\
633 & 1.22 & -0.01 & 0 \\
1175 & 1.43 & -0.18 & 0 \\
1225 & 0.21 & 1.03 & 0 \\
1314 & 0 & 0 & 0 \\
1418 & 0 & 0 & 0.01 \\
1592 & 0 & 0 & 0 \\
1682 & -0.01 & -0.22 & 0 \\
\hline
\end{tabular}
\caption{The table lists energies and transition amplitudes for crystal field and spin-orbit states in an exchange field of 250 meV.}
\end{table}

Since a mean exchange field of 250 meV has been included, the results of Fig. \ref{cef_2} simulate the single-ion spectrum of the ordered antiferromagnet. The calculation with large exchange field along the $z$ direction shows three peaks in the energy range of interest while the neutron spectrum of paramagnetic YBCO shows one, one that correponds to the calculated out-of-plane $zz$ fluctuations.  Existing measurements of antiferromagnetically ordered YBCO and LSCO have not to date detected such structure in the $\sim$ 400 meV region, a difficult region for neutron spectroscopy.  However, the calculated fluctuations have energies in doped YBCO comparable to our experiment.

Table III shows the amplitude of the transition magnetic momenta, $\vec{M}=\vec{L}+2\vec{S}$, whose square gives the magnetic peak intensity arising from transverse excitations of $M^{+}$ and $M^{-}$ and from out of plane fluctuations $M^{z}$, which is longitudinal for exchange field in the z-direction.

The ``spin-wave" peak in Fig. \ref{cef_2} is the lowest energy peak and its intensity is proportional to the square of the transverse spin matrix element $M^{x}=(M^{+}+M^{-})/2$.  Only M$^{-}$=1.89 (Table III) connects the ground state to the first excited state making the spin-wave transition strength $(M^{x})^{2}$=0.89 while the longitudinal orbital transition at 390 meV has a greater strength $(M^{z})^{2}$=2.99.

	Our calculations indicate that direct magnetic dipole transitions occur in the energy range accessible to neutron and x ray inelastic scattering and should not be ignored.   This magnetic spectrum represents the insulating background upon which changes are made by hole doping.  Its spectrum is physically distinct from and of higher energy than the strong attenuation of magnetic spectral weight we have observed as the spin waves enter a pseudogap continuum above 250 meV.   Some x ray papers have suggested (for example, Ref.\onlinecite{Braicovich09:102,Hill08:100}) that relatively low energies less than $\sim$ 1 eV cannot arise from $dd$ transitions.  Our relatively simple estimates of spin and orbital magnetism suggest otherwise. The orbital moment is not quenched since it contributes half the ground state moment, adds a substantial xy symmetry  into the ground state, and gives a large intensity for inter-orbital transitions.  We have therefore shown that orbital moments play an important role in the energy spectrum when transitions out of the ground state are made at a few hundred meV.

\end{document}